\documentclass[12pt,journal,onecolumn,draftcls,peerreview]{IEEEtran}
\usepackage{amssymb}
\usepackage{amsmath}
\usepackage{amsfonts}
\usepackage{graphicx}
\usepackage{algorithm}
\usepackage{algorithmic}
\usepackage{cite}
\usepackage{epstopdf}
\usepackage{multirow}
\usepackage{stfloats}

\newtheorem{definition}{\textbf{Definition}}

\newtheorem{lemma}{\textbf{Lemma}}

\newtheorem{proposition}{\textbf{Proposition}}

\begin{document}

\title{On Content-centric Wireless Delivery Networks}
\author{Hui Liu, ~\IEEEmembership{Fellow,~IEEE,} Zhiyong Chen, ~%
\IEEEmembership{Member,~IEEE,} Xiaohua Tian, ~\IEEEmembership{Member,~IEEE,}
Xinbing Wang, ~\IEEEmembership{Senior Member,~IEEE,} and Meixia Tao, ~%
\IEEEmembership{Senior Member,~IEEE} \thanks{%
The authors are with the Institute of Wireless Communications Technology
(IWCT), Department of Electronic Engineering, Shanghai Jiao Tong University,
Shanghai, 200240, P. R. China. Emails: \texttt{\{huiliu, zhiyongchen, xtian,
xwang8, mxtao\}@sjtu.edu.cn}}}
\maketitle

\begin{abstract}
The flux of social media and the convenience of mobile connectivity has
created a mobile data phenomenon that is expected to overwhelm
the mobile cellular networks in the foreseeable future. Despite the advent
of 4G/LTE, the growth rate of wireless data has far exceeded the capacity
increase of the mobile networks. A fundamentally new design paradigm is
required to tackle the ever-growing wireless data challenge.

In this article, we investigate the problem of massive content delivery over
wireless networks and present a systematic view on content-centric
network design and its underlying challenges. Towards this end, we first
review some of the recent advancements in Information Centric Networking
(ICN) which provides the basis on how media contents can be labeled,
distributed, and placed across the networks. We then formulate the content
delivery task into a \emph{content rate} maximization problem over a share
wireless channel, which, contrasting the conventional wisdom that attempts
to increase the bit-rate of a unicast system, maximizes the content delivery
capability with a fixed amount of wireless resources. This conceptually
simple change enables us to exploit the \textquotedblleft content
diversity\textquotedblright\ and the \textquotedblleft network
diversity\textquotedblright\ by leveraging the abundant computation sources
(through application-layer encoding, pushing and caching, etc.) within the
existing wireless \ networks. A network architecture that enables wireless
network crowdsourcing for content delivery is then described, followed by an
exemplary campus wireless network that encompasses the above concepts.
\end{abstract}

\newpage

\section{Background and motivations}

The growing popularity of smart mobile devices, coupled with bandwidth
hogging data services and applications (e.g., Youtube), has spurred the
dramatic increase in mobile traffic volumes. It is widely anticipated that
the amount of mobile traffic beyond 2020 will be 1000 times higher than the
2010 traffic level \cite{cisco_1000}. This so-called \textquotedblleft
mobile data tsunami\textquotedblright\ has driven the mobile operators and
wireless researchers around the world to the edge: how can the wireless
infrastructures and radio communication technologies advance fast enough to
keep up with this phenomenon? Various innovative capacity-increasing
solutions have been proposed and investigated.
Among other promising techniques, i) denser base station deployment (by a
factor of 10 in areas with a large density of active users), ii) additional
spectra, and iii) improved spectral efficiency (by a factor of 10 through
distributed antenna systems (DAS), cell collaborations, etc.) are widely regarded as the cornerstones technologies for next
generation mobile networks.

\subsection{The myths of exponential increase in mobile traffics}

While the collective solutions may indeed deliver the 1000 times capacity
increase within this decade, they do not adequately address two of the
fundamental issues associated with mobile traffic explosion.

\begin{itemize}
\item Sustainability and scalability: Given the Shannon theory bounded radio
link capability, the scarcity of mobile spectra, and the environmental and
operational constraints, it is unlikely that the current wireless bottleneck
can be averted solely by wireless infrastructure expansions.

\item Mobile traffic characteristics: Media contents constitute the lion's
share of today's mobile traffic\footnote{%
Given the stochastic nature of wireless data, the 1000 times traffic volume
increase itself does not necessarily mandate a 1000 times increase in mobile
cellular capacity, defined as bits/second/Hz/m$^{2}$.}. More importantly,
the traffic characteristics have changed profoundly over time \cite{Media}.
As such, it becomes questionable whether the unicasting optimized cellular
networks are the right platform for the task of massive content delivery.
\end{itemize}

The first issue is easily appreciated since the mobile traffic growth is
unlikely to stop by 2020. With diminishing returns in infrastructure-based
investments, it is difficult to imagine that the next wave for mobile data
tsunami can still be tackled with infrastructure and bandwidth expansions.
The second issue, on the other hand, offers both challenges and promises at
the same time. Upon a closer examination at the wireless traffic trend, it
is observed that despite of their massive numbers, the amount of
\textquotedblleft unique\textquotedblright\ media content does not actually
increase exponentially \cite{Fetocache_magazine}. In addition, even unique
multimedia contents are not consumed with the same frequency, at least
statistically. Studies have revealed that

\begin{itemize}
\item 70\% of the wireless traffic is from videos \cite{cisco_1000}, many
with applications that personalizes the viewing experience. Specifically,
users expect to be able to watch recently broadcast content and also access
archived programs across different mobile platforms..

\item Only a small percentage, 5-10\%, of \textquotedblleft
popular\textquotedblright\ contents are consumed by the majority of the
mobile users, despite the large temporal variability in the content
consumption time. In particular, \textquotedblleft
personalized\textquotedblright\ viewing experience can be achieved through
content pushing/caching and time-shifting \cite{Media}.

\item Modern media contents are increasingly decoupled from their sources.
The contents of today can be modularized, labeled, separately delivered and
placed across different nodes in the network, and reassembled at the user
end at the time of request \cite{Media}.
\end{itemize}

The above observations suggest that mobile applications are undergoing a
fundamental shift, from the conventional \textquotedblleft
connection-centric\textquotedblright\ behaviors (e.g., phone calls, text
messages), to a more \textquotedblleft content-centric\textquotedblright\
usage model. The cellular networks of today, which utilize technologies such
as micro- and pico-cells, frequency reuse, and directional transmissions,
etc., are optimized for unicasting services so that unique information can
be delivered to individual users. However, it can be argued that the logjam
of the cellular networks is not due to the lack of \textquotedblleft
connection\textquotedblright\ capability (i.e., unicasting). Instead, the
real problem is the ineffectiveness of the current cellular architecture in
massive content delivery. In fact, it has been suggested that a converged
broadcast and cellular network, akin to an integrated high-speed train and
highway transportation system, might be the most efficient combination for
wireless media content delivery.

\subsection{From connection-centric to content centric}

Before we dive deep into the content-centric wireless network design, a
quick overview of the upper-layer advances in content creation and
distribution is due first.

The inability of the mobile networks to sufficiently exploit the content
characteristics is in part attributable to how contents are created and
accessed in the IP-based network. Traditionally, mobile content accessing is
achieved by establishing a \emph{link session} between the content host and
the client via wired-line and wireless infrastructures. A majority of these
link sessions are IP-based. The current Internet architecture was designed
under the host-centric communication paradigm, which is problematic in
meeting the demands of scalable multimedia applications \cite{icn13}.

\begin{figure}[t]
\centering
\includegraphics[width=5in]{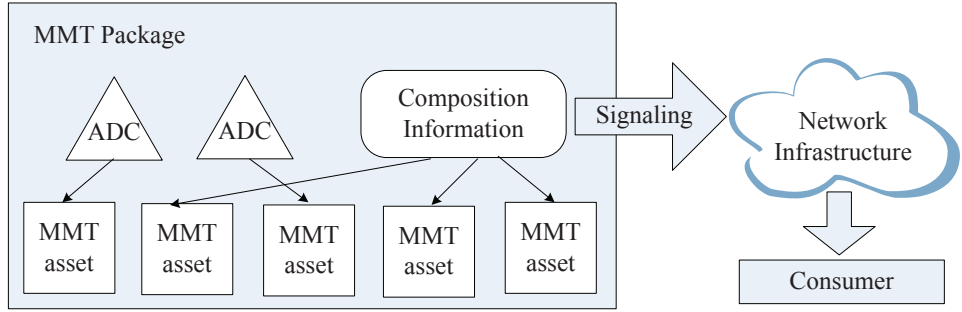}
\caption{MMT assets and service packages.}
\label{MMT_figure}
\end{figure}

The need of fully utilizing the network resources for content distribution
has motivated the development of future network architectures based on the
named data objects (NDOs), which is commonly known as the
information-centric networking (ICN) \cite{icn13}. A fundamental
feature that distinguishes the ICN model from the current Internet is that
the information/content is decoupled from its sources, so that a particular
content can be named and placed anywhere within the network. This feature
not only facilitates the information distribution, but also provides the
freedom in the creation of personalized, on-demand multimedia contents. For
example, a popular video could be embedded with personalized advertisement
tailored to the user's preference, and patched with audio in a language
suitable for users in an area. In this scenario, each of the components or
the \emph{content objects}, could be assigned with a name, delivered through a
different path, and be combined towards the needs of different users.

The MPEG in particular is developing the MPEG Media Transport (MMT) standard
in order to enable flexible content access with uniquely identifiable names
for optimized content delivery \cite{mmt}. The MMT package is a logical
entity consisting of MMT Assets, Composition Information (CI), and Asset
Delivery Characteristics (ADCs), which are coded media data about the
content, the information about how the data should be combined and
delivered, respectively - see Fig. \ref{MMT_figure}. Compared with the
connection-based delivery, where the service package is delivered in its
entirety, the NDO content structure provides much higher flexibility and
significant saving in network resources, and lands itself nicely into a
content-centric wireless framework to be discussed in the ensuring sections.

In the rest of the article, we shall seek solutions to the wireless
bottleneck problem by taking advantages of the newly equipped NDO features
and the readily available computation resources (e.g., application-level
encoding and caching) within the wireless network. The goal is to arrive at
a content centric design that could potentially offer multiplicatively
improvement. Specifically, we shall focus on maximizing the content delivery
capacity \emph{with a fixed amount of wireless resources}. With this purpose
in mind, we organize the remainder of the article is as follows.

\begin{itemize}
\item In Section II, we describe the new dimensions that could be exploited
to achieve the set goals and provide a high level overview on the state of
the art.

\item In Section III, we formulate wireless content delivery into a \emph{%
content rate} maximization problem over a shared wireless channel. This
change of design paradigm enables us to quantify the \emph{content diversity}
in wireless delivery. A priority encoding transmission (PET) \cite{PET}
based\ joint transmission scheme is then described, which reveals important
tradeoffs on how computations (encoding and caching) can affect the
wireless content delivery capability.

\item In Section IV, we focus on the \emph{network diversity} and generalize
the framework to multiple wireless networks where the notion of
\textquotedblleft network crowdsourcing\textquotedblright\ for content
delivery is established. A network architecture that enables wireless
network crowdsourcing is then described.

\item Finally in Section V, we illustrate the feasibility of content-centric
wireless networks with an exemplary campus wireless network that encompasses
the above concepts.
\end{itemize}

\section{Exploiting new dimensions in wireless content delivery}

Historically, leapfrogs in mobile communications are often associated with
the discovery of new dimensions and diversities. The breakthroughs in
time-frequency, code, space, and multiuser diversities have been the sources
of wireless bit rate breakthroughs (by three orders of magnitude) for the
past three decades. In light of the fact that the mobile media contents are
highly diversified, both in characteristics and distribution channels, we
can anticipate a dramatic increase of the ability of mobile content delivery
by exploiting the so-called \textquotedblleft content
diversity\textquotedblright\ (at application level) and \textquotedblleft
network diversity\textquotedblright\ (at infrastructure level), via
approaches that take advantage of the computation and caching capabilities
within the wireless networks.

Various approaches have been attempted to exploit the content and wireless
network diversities. In particular, the so-called \emph{converged network},
which combines broadcasting with cellular, provides a highly efficient and
practical means to simultaneously capture the content and network diversity
gains. A number of results have been reported on converged network. The
authors in \cite{adaptiveScheme} proposed and analyzed a hybrid scheduling
scheme for mixed multicast and unicast traffics in cellular systems. A
collaborative scheme is proposed in \cite{cost_converged} to optimize the
trade-off between the amount of parity data transmitted by the broadcasting
station and repair data delivered through cellular channels to users with
missing information from broadcasting. For the purpose of reducing the
delivery cost, the authors in \cite{rrm_converged} adopt application layer
forward error correction with Raptor coding to repair errors of the initial
DVB-H broadcast using HSDPA and MBMS channels. The converged
information delivery was considered in 3GPP and 3GPP2 back in 2002, when
both organizations created items for broadcast/multicast services in
GSM/WCDMA and CDMA2000, respectively. Broadcast services delivered in 3G
networks is introduced in \cite{deliverBroadcast3G}, in which the authors
argue that the hybrid unicast-broadcast delivery is superior not only in
terms of the system resource usage but the user experience. Most recently, $AT\&T$ announced its
plan to use $700$Mhz channels for LTE Broadcast networks to
remove video from its wireless network, clearing those airwaves up for other
data.

At the algorithmic level, the content diversity gains are achieved by utilizing the computation power (e.g., application layer
encoding and decoding) and caching capacities within the wireless networks.
The major benefits of content caching in wireless applications include i)
minimizing the content downloading time, ii) alleviating the traffic loads
on the core networks, and iii) mitigating the over-the-air wireless
traffics. Niesen et al investigated the inner- and outer- bounds on the
caching capacity region for a generic wireless network with $n$ nodes \cite%
{capacity_cache}. Most practical caching schemes assume a \emph{two-phase
caching strategy} which involves a placement phase and a delivery phase \cite%
{fund_caching}. By delivering the popular contents, either in their entirety
or in pieces, to nearby stations (e.g., basestations, relay nodes, helper
nodes, etc.) during the placement phase, the user \emph{downloading delay
time} can be minimized \cite{Fetocache_TIT}. These approaches however,
do not actually reduce the wireless traffics within the networks.

For the purpose of minimizing the over-the-air wireless traffics, it has been
proposed that the contents be placed to user devices during the wireless
off-peak time, so that the wireless network peak rate can be greatly reduced
\cite{fund_caching}. A novel architecture with distributed caching of video
contents in femto-base stations and wireless terminals is described in \cite%
{Fetocache_magazine}. Several network coding-based content placement
approaches have been proposed and analyzed \cite%
{fund_caching}.  All these approaches assume a high level of \emph{temporal
variability} in wireless network traffic. On the other hand, the two-phased approaches do not reduce
the total amount of the over-the-air traffic loads. To date, only limited
research attempts to combine the content diversity with caching/encoding in
order to reduce the \emph{total amount of wireless traffics}.

\section{Content-centric wireless delivery}

Despite the various schemes that exploit the content characteristics in
wireless delivery, the \emph{content diversity} has not been quantified
under a unified framework, even under the simplest multiuser scenario. In
this section, we introduce a novel information-theoretical formulation for
the content delivery problem. In particular, the notion of wireless \emph{%
content rate} is established, which, in contrast to the \emph{bit\ rate} of
a wireless unicast system, characterizes a direct relation between the
content diversity and the content delivery capability of a wireless system.

\subsection{Content rate\ formulation}

The capacity of a wireless system is traditionally measured by the bit rate
at which unique information is reliably transmitted over the wireless
system. In order to quantify the effectiveness of wireless content delivery,
we must first distinguish the bit-rate with the so-called \emph{content rate}%
, defined in the ensuring section. We will show that for a wireless system
with \emph{fixed bit-rate}, its content rate could be increased
substantially, depending on the content diversity and the computation power
within the wireless system.

\begin{figure}[t]
\centering
\includegraphics[width=5in]{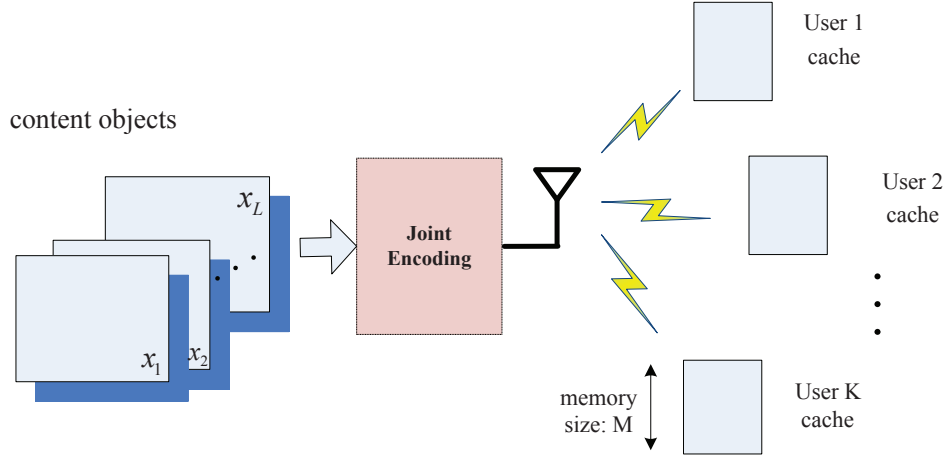}
\caption{A scenario where $L$ content objects are available for delivery to $%
K$ users in the wireless system.}
\label{system_model}
\end{figure}

To elaborate, consider the scenario in Fig. \ref{system_model} where $L$
content \emph{objects} (e.g., MMT assets) are available for delivery to $K$
users in the wireless system. The set of independent content objects is
denoted as $\{x_{l}\}$, each with the size $|x_{l}|$ bits, $l=1,\cdots ,L$.
To receive personalized services, each user would request a particular
service package at a specific time. We denote such a request from the $k$-th
user at time $t_{k}$ as $y_{k}(t_{k})$. Unlike a conventional wireless
system which delivers $y_{k}$ in its entirety through unicasting, portions
of the content objects will be sent and cached at the user end before the
actual request. To understand the above process, let
\begin{equation}
y_{k}(t_{k})=\left\{ f_{k}(x_{k_{1}},x_{k_{2}},\cdots
,x_{k_{i}}),t_{k}\right\} ,
\end{equation}%
where $(x_{k_{1}},x_{k_{2}},\cdots ,x_{k_{i}})$ are the set of content
objects needed for request $y_{k}$, and $f_{k}$ is the content
representation function, i.e., how content objects are processed and
represented to the user. For simplicity, we shall assume $f_{k}=\bigcup
\left\{ x_{k_{1}},x_{k_{2}},\cdots ,x_{k_{i}}\right\} $, i.e., the union of
all the content objects needed by $k$-th user in the remainder of this
paper.

Clearly, as long as the corresponding content objects are received by the
user before $t_{k}$, the $k$-th user's request can be satisfied.
Accordingly, we can define the wireless content rate as \emph{the rate at
which the amount of service packages }$\{y_{k}(t_{k})\}$\emph{ is successfully
delivered to users over the shared wireless channel.} Such a measure is
more suitable for the problem of interest since, for content delivery
purpose, the users are only concerned with the amount of service packages
received while the wireless carriers are mostly keen on the number of
satisfied users on the network. To formulate the content rate
mathematically, assume a \emph{shared} wireless channel with fixed link rate
$=1$b/s/Hz. Further denote

\begin{itemize}
\item $B$: the bandwidth of the wireless channel

\item $M$: the maximum user cache length

\item $\mathbf{Z}$: the content diversity matrix ($K\times L$) that maps
content objects to users. Each element of the matrix is either 1 or 0. If $%
Z_{kl}=1$, it means the $k$-th user is interested in the $l$-th content
module; otherwise it is not.
\end{itemize}

\begin{definition}
Given a fixed amount of wireless resource $B\times T$ ($T=\max \left\{ {t_{l}%
}\right\} $), the \emph{content rate} $R_{C}$ is defined as the total amount
of service packages successfully delivered to the users:
\begin{equation}
R_{C}=\frac{\sum_{k=1}^{K}|y_{k}(t_{k})|}{B\times T}.
\end{equation}

\begin{definition}
$\{R_{C},M,\mathbf{Z}\}$ is \emph{achievable} if there exists a transmission
strategy for a given wireless resource $B\times T$ such that all users are
able to successfully receive their requested service packages before their
respective requesting time instants.
\end{definition}
\end{definition}

\bigskip

Since $R_{C}$ is inverse proportional to the bandwidth $B$, the content rate
is maximized when all users requests are met with the minimum amount of
bandwidth: $\min B\Longrightarrow \max R_{C}$.

\subsection{\protect\bigskip The content rate bounds}
Determining the achievable $\{R_{C},M,\mathbf{Z}\}$ is non-trivial to say
the least. First of all, the content objects $\{x_{l}\}$ are neither \emph{%
private} nor \emph{common}, and therefore, they cannot be efficiently
delivered with a unicasting or a broadcasting mechanism alone. The content
delivery ability of the wireless system depends on several factors:

\begin{enumerate}
\item how the content objects are mapped to the user requests (i.e., $%
\mathbf{Z}$);

\item how content objects are jointly encoded and delivered over the shared
wireless channel;

\item the amount of caching available at the user ends, so that content
objects can be delivered before the service requests.
\end{enumerate}

Interestingly, the later two factors are determined by the \emph{computing
capability}, i.e., application-layer encoding/decoding and caching, rather
than the communication capability of the wireless system. As such,
optimizing the content rate must be conducted under a \emph{%
computing-communication} framework, which simultaneously exploits the
communication and computation limits of the wireless system.

The following lemmas shed some light on how the content rate relates to the
wireless system bit rate.

\begin{lemma}
The content rate \emph{upper bound} of a shared wireless channel is reached
when $\left\{ t_{i}\right\} =T$, $M=\infty $ (i.e., infinite caching), and
all content objects being broadcast and cached at the user ends. In this
case, the minimum amount of bandwidth required is given by $B_{\min
}=\sum_{l=1}^{L}|x_{l}|/T$.
\end{lemma}

\begin{lemma}
The content rate \emph{lower bound} of a shared wireless channel is reached
when\emph{\ }$M=0$ (i.e., zero caching), and each content object being
unicasted to individual users at different requesting times. In this case,
the maximum amount of bandwidth required is given by $B_{\max
}=\sum_{k=1}^{K}|y_{k}|/T$, and the content rate reduces to $R_{c}=1$b/s/Hz,
the wireless system bit-rate.
\end{lemma}

\bigskip

In other words, broadcasting with unlimited caching at the user ends
delivers the highest content rate, whereas unicasting upon individual
service request is the least efficient way of using the wireless resources
in content delivery.
\begin{figure}[t]
\centering
\includegraphics[width=5in]{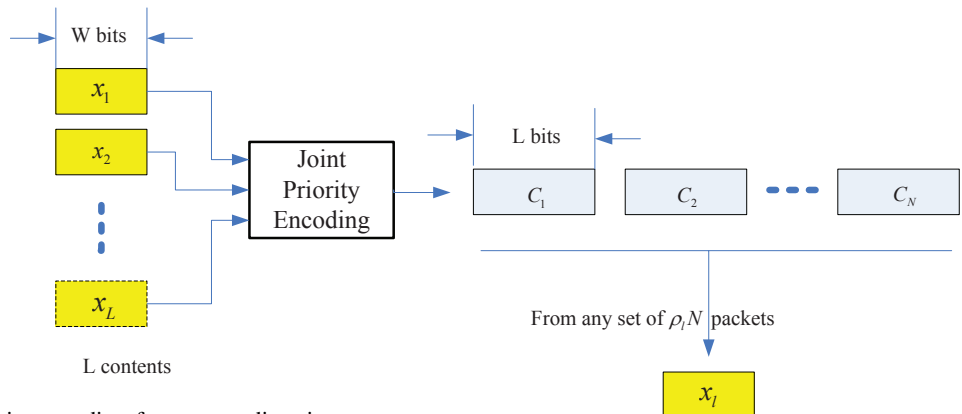}
\vspace{-1cm}
\caption{Priority encoding for content diversity.}
\label{priority_BC}
\end{figure}

\subsection{Priority encoding for content diversity}
In practice, approaches such as application-layer encoding with user-end
caching can be employed to increase the content rate $R_{c}$. An elegant
scheme, termed the \emph{priority encoding transmission} (PET), was
introduced by Albanese et at in 1996 \cite{PET} to capture the content
diversity with joint source encoding.

Consider the scenario depicted in Fig. \ref{priority_BC}, where the content
objects $\left\{ x_{l}\right\} $ are to be delivered to the users with
caching capability. Instead of sending the objects individually, the content
objects can be jointly encoded to save the wireless bandwidth. At the same
time, since the content objects are not requested at the same frequency, the
content priorities must be accounted for at the encoding stage to benefit decoding. Applying the PET principle to wireless content delivery, $%
\left\{x_{l}\right\} $ are jointly encoded with \textit{priority} based on
their\textit{\ }degree-of-interests.

\begin{proposition}
Let $\rho _{l}\subset (0,1],l=1,...,L$, be the priority index of \textit{the
}$l$\textit{th content object}. \textit{There exists a priority encoded
sequence consists of (Albanese et al \cite{PET}): An encoding function that
maps the content set }$\left\{x_{l}\right\} $\textit{\ onto an encoded
sequence of }$N$\textit{\ packets, with }$\Gamma $\textit{\ bits each, and a
decoding function that maps the }$N$\textit{\ packets onto the }$L\ $\textit{%
contents. The decoding function is able to decode the }$l$\textit{-th content
from any }$\rho _{l}$\textit{\ fraction of the }$N$\textit{\ encoding
packets. }
\end{proposition}

By making $\{\rho _{l}\}$ inverse proportional to the content popularity,
content objects with higher priorities will be decodable from fewer coded
packets, whereas objects with lower priorities are decodable from more
packets. Applying the PET scheme to jointly content object encoding as shown
in Fig. \ref{priority_BC}, the content rate is increased while the average
caching size is also minimized accordingly.
\section{Wireless network CROWDSOURCING}
Given that the modern contents can be distributed and placed across the
networks, and the fact that wireless users often have access to various
forms of wireless resources (i.e., network diversity), we examine in this
section the feasibility of wireless content delivery over multiple wireless
infrastructures, potentially from different wireless network service
providers (NSPs).

The current wireless NSPs face multifaceted challenges. On one hand, the
amount of traffics delivered over their wireless networks will continue to
grow exponentially in the foreseeable future. On the other hand, the profits
made from traditional pipe-like services are not increasing in line with the
traffic increase rate. Particularly, the super active over-the-top (OTT)
service providers (e.g., Google, Amazon, Netflix, etc) are revolutionizing
the way people are using the network, and at the same time, undercutting the
traditional business model of the NSPs. The most representative OTT service
is the high-quality multimedia distribution, which is and will be the
dominant wireless traffics. The accelerating trend is motivating NSPs to
seek innovative service models and partnership in responsive to the new
economy driven by the OTT innovation.
\begin{figure}[t]
\centering
\includegraphics[width=4in]{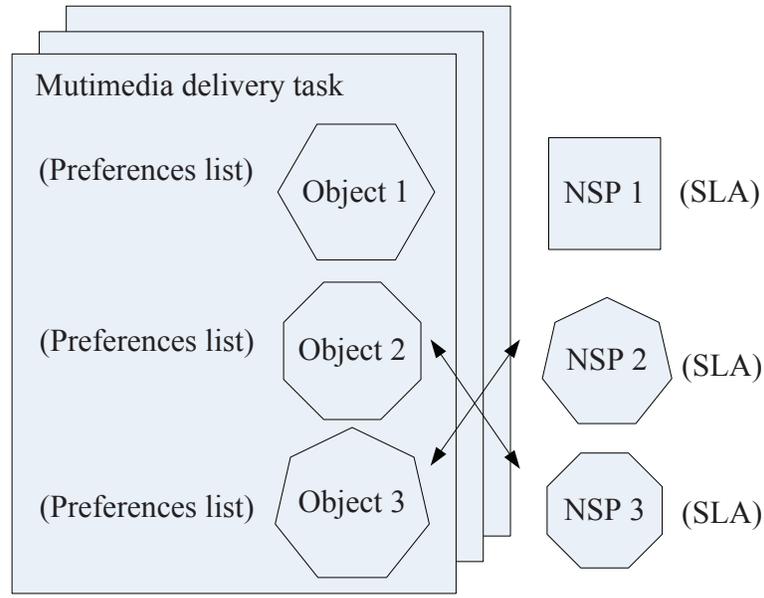}
\caption{Network crowdsourcing for wireless content distribution.}
\label{matching}
\end{figure}

We propose a collaborative model between the OTT Content Provider (OCP) and
wireless NSPs termed as \emph{Network Crowdsourcing}, which is illustrated
in Fig. \ref{matching}. The essence of Network Crowdsourcing is to distribute content objects according to OCP's requirements by soliciting distribution services from a group of NSPs,
rather than from a unique NSP as in the traditional way. The NSPs are acting as contractors to undertake the
task generated by the OCP, where each task should be the delivery of
modularized content objects. Each NSP provides the OCP a service level
agreement (SLA) showing how certain object could be delivered. The SLA
should specify at least the following information: i) the network resources
to be allocated to the object distribution, ii) the expense the OCP should
spend for the object distribution, and, iii) the management functionalities
can be provided to the OCP. Based on the SLAs offered by NSPs, the OCP may
choose a group of NSPs to accomplish the content distribution task,
therefore the notion of ``network crowdsourcing". The OCP can also provide
task profiles specifying the preferences of each objects distribution
subtask, including the amount of resource needed, budget and preferred
management interfaces.

Under the network crowdsourcing model, the OCP and clients could communicate
with each other on which kind of network resources the client is able to
utilize. For example, the client user may see multiple wireless networks
owned by different NSPs, and each NSP may provide different kinds of access
schemes such as broadcasting, WiFi and cellular. By aggregating such
information, the OCP could understand the popularity of each object and the
geographic distribution of clients served by each NSP, which could \ in turn
greatly help the OCP generate the task profile.

From the perspective of the NSP, the current trend towards software defined
networking (SDN) \cite{mobileflow13} technology provides the unique
opportunity for the NSP to obtain more effective control and scheduling of
its network resources. In particular, the SDN decouples the control plane
and data plane of networking routers and switches, which enables the global
control and management of the network fabric. The SDN controller can
facilitate checking whether the required network resources are available;
moreover, it can provide interfaces for applications to realize certain
control and management functionalities.
\subsection{Network Architecture and delivery schemes}

\begin{figure}[t]
\centering
\includegraphics[width=5.7in]{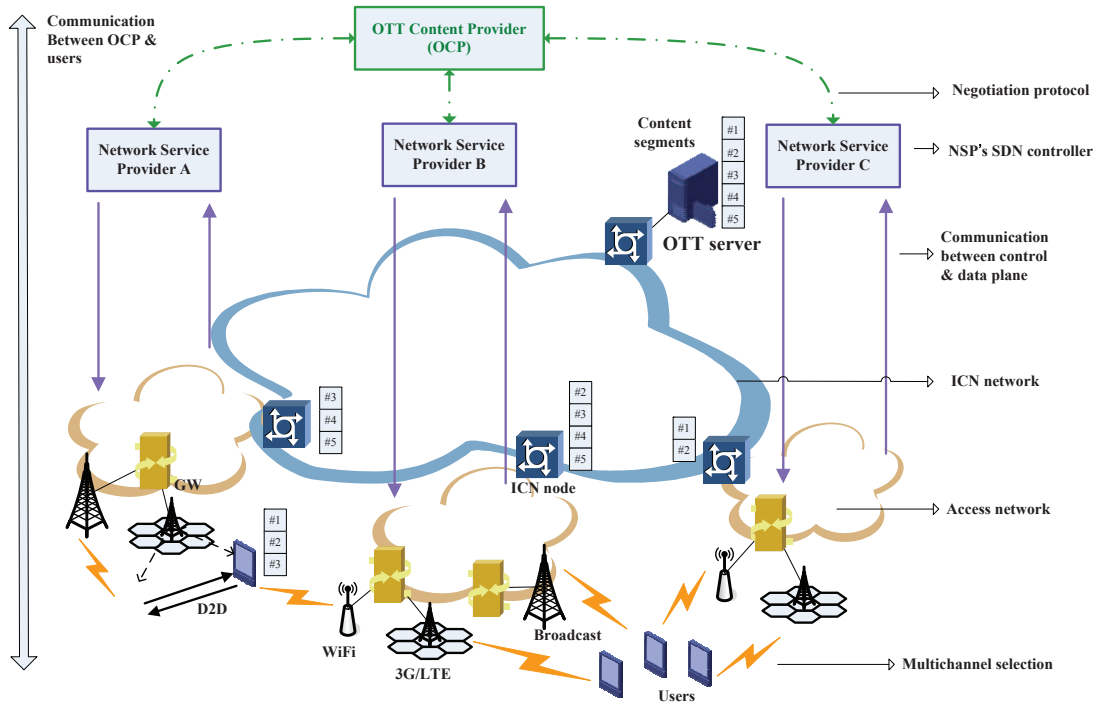}
\caption{A network architecture that enables network crowdsourcing in
massive content delivery.}
\label{arch}
\end{figure}

Under the above framework, we now describe a network architecture that
enables network crowdsourcing in massive content delivery, as shown in Fig.~%
\ref{arch}. The architecture consists of three entities: OCP, NSPs and users. The OCP wants to distribute its modularized objects to users with the services of NSPs, with the users' context and popularity of each object taken into account. Specifically with this architecture, the OCP can negotiate with
the NSPs for distributing contents, based on the task preference and SLA
provided by NSPs. Through interactions, the OCP develops an understanding of
the content demand distribution and the particular users' contexts through
its communication with users. Each NSP maintains effective control over its
networking resources, with the facilitation of SDN or similar technologies.
The negotiation protocol can work on the application layer interfacing the
OCP and NSP's network management application over the controller. After the
negotiation process, the network distributes the modularized objects over
the ICN. According to the popularity of each object and the location
distribution of users, the objects will be pushed to and cached at
corresponding parts of ICN, in order to facilitate wireless delivery through
access networks. Users will obtain the requested content through channels
based on different access techniques, and even from different NSPs. For example, the OCP needs to distribute 5 kinds of modularized objects in its server to users, as shown in Fig.5. With the negotiation and distribution process described above. Each object is cached at different locations with different amounts of copies, which complies users' geographic distribution and their demands of each object. The
objects can then be reassembled to form a playable content.

\section{\protect\bigskip An exemplary converged network}

In this section, we present an exemplary wireless campus network where the
content diversity and the network diversity is exploited within a
broadcast-cellular-WiFi converged network.

Lemma 1 in Section III-B suggests that with sufficient caching capability at
the user ends, broadcasting may be the most efficient means to deliver
massive contents to wireless users. In practice however, the memory size at
a mobile device is always limited. In addition, the long tail of the Zipf
distribution cannot all be accommodated with broadcasting. A
broadcast-cellular converged network provides a more sensible solution to
the wireless content delivery \cite%
{adaptiveScheme,cost_converged,GC_Conver}.

\subsection{The large-scale campus wireless network}
\begin{figure}[t]
\centering
\includegraphics[width=5.7in]{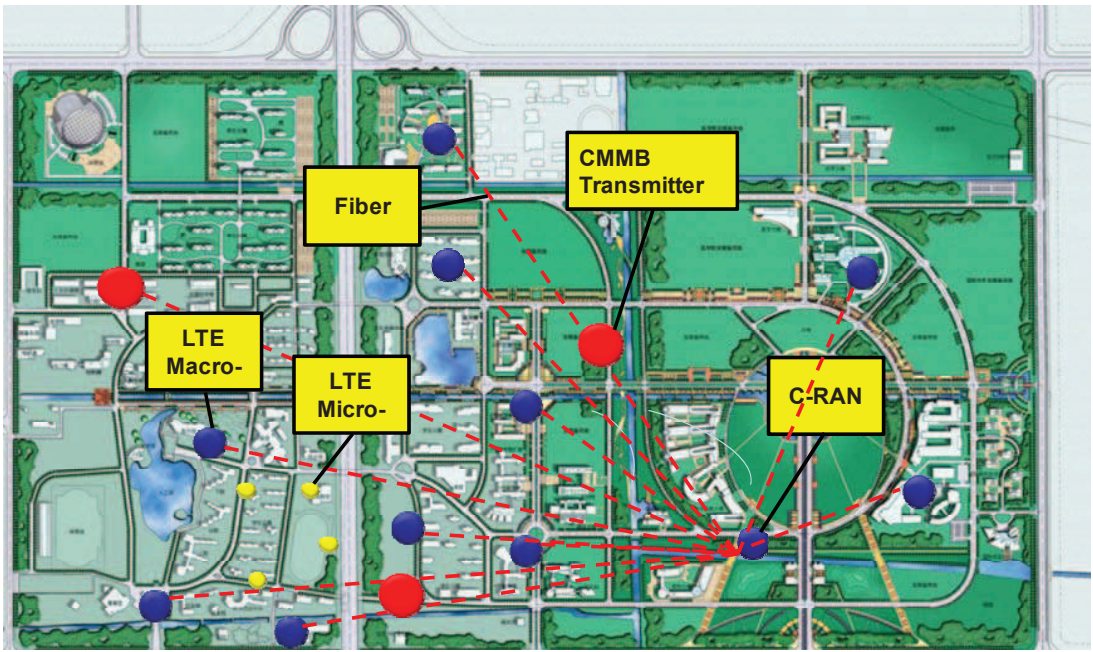}
\caption{A large-scale wireless campus network in Shanghai Jiao Tong
University}
\label{campus}
\end{figure}

In collaboration with Huawei, a large-scale wireless test network has been
established on the 3 square kilometer campus of Shanghai Jiao Tong
University (SJTU). The deployment is scenario depicted in Fig. \ref{campus}.
The heterogenous wireless network is comprised of an LTE cellular system, a
CMMB based digital broadcasting system, and a WiFi system. In particular,
the campus network features

\begin{itemize}
\item 80+ micro- and pico- stations, providing blanket coverage of the
entire campus

\item over 66 kilometers of fiber, connecting all cellular RF transceivers
to form a truly cloud-based radio access network (C-RAN)

\item 3 single-frequency-network (SFN) digital broadcasting stations

\item 2500+ WiFi APs across the campus
\end{itemize}

The broadcasting system, the cellular system, and the WiFi system work
collaboratively provide a high-speed data and media delivery services to
trial mobile users/clients that include students and faculty, campus
shuttles, wireless surveillance cameras, EVs, etc. The users under the
coverage of the converged network can receive services from both the
broadcasting system and the cellular/WiFi system.

\begin{figure}[t]
\centering
\includegraphics[width=4in,height=3.7in]{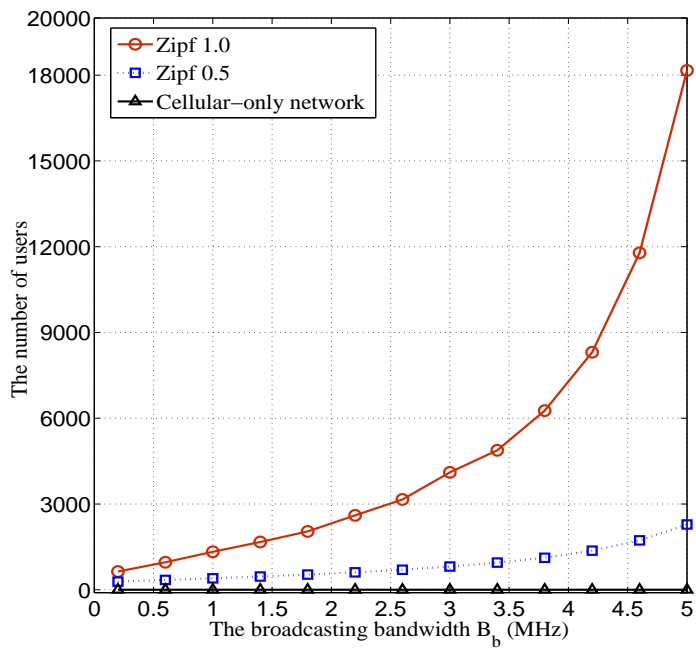}
\caption{The number of users supported in a cell of the converged network
and the cellular-only network.}
\label{capacity}
\end{figure}

Specifically for the content delivery applications, the three wireless
systems are \emph{converged to form a push-based service platform, }so that
the multimedia contents can be delivered to the users via: i) broadcasting
with pushing and caching, and/or ii) unicast delivery through cellular/WiFi
networks. A \emph{content scheduler} is utilized to update the pushing list
which is comprised of the most popular contents \cite{GC_Conver}.  In Fig.\ref{capacity}, the number of users per cell supported in the push-based converged network is simulated. Here, the Zipf distribution
is used to characterize the content popularity distribution, in
which the Zipf parameter (e.g., 0.5 or 1) determines how diverse the contents are. The popularity would become
more skewed with the larger Zipf parameter. The results
clearly show that the push-based converged network can
bring potentially \emph{multifold} benefits in terms of content delivery
capacity as we moderately increase the broadcasting bandwidth. We also see that the more ``concentrated'' the contents are, the more gains the push-based converged network can provide.

Currently, technical trials that involves about 500 users are underway on
the campus converged network. The goal is to implement selective algorithms
and schemes, e.g., the PET scheme and the network crowdsourcing scheme
described in this article, so that they can be validated and optimized for
real world applications.
\section{\protect\bigskip Concluding remarks}
In this article, we have exploited and motivated a content-centric design
framework for wireless massive content delivery. The centerpieces of our
discussions are the application-level \emph{content diversity} and the
infrastructure-level \emph{network diversity}, which we assert can be
effectively captured by taking advantages of the computation resources
within the wireless networks. For the content diversity specifically, we
have formulated the \emph{content rate} to provide a precise
measure on the content delivery efficacy of a wireless system. A
corresponding joint encoding scheme based on the PET algorithm has been
presented. On the network diversity side, we have introduced the notion of
\emph{network crowdsourcing }and its supporting architecture to enable
content delivery across different wireless platforms. As part of the ongoing
and future work, we are implementing and evaluating selective design
concepts on a large-scale wireless test network on the SJTU campus. We
anticipate the outputs to provide validations and critical insights on the
design of content-centric wireless delivery networks.

\section{\protect\bigskip ACKNOWLEDGMENT}
This work is supported in part by the the National High Technology Research and Development Program of China under 863 5G Grant 2014AA01A702, the National Natural Science Foundation of China (61221001, 61301115, 61322102, 61202373), the China Postdoctoral Science Foundation (2012M520897, 2014T70416), the Shanghai International Cooperation Project (13510711300), and Shanghai Basic Research Key Project (No.13510711300, 12JC1405200, 11JC1405100).

\bibliographystyle{IEEEtran}
\bibliography{content_journal}
\begin{IEEEbiography}{Hui Liu}
 is the Zhi Yuan Chair Professor of Shanghai Jiao Tong University, received his B.S. in 1988 from Fudan University, Shanghai, China, and a Ph.D. degree in 1995 from the Univ. of Texas at Austin, all in electrical engineering. He was previously an assistant professor at the Dept. of EE at Univ. of Virginia and a full professor at the Dept. of EE, Univ. of Washington. Dr. Liu was the chief scientist at Cwill Telecom, Inc., and was one of the principal designers of the TD-SCDMA technologies. He founded Adaptix in 2000 and pioneered the development of OFDMA-based mobile broadband networks (mobile WiMAX and 3G LTE). Dr. Liu is the creator of CMMB transmission technology which enables the delivery of mobile TV services to more than a billion of population in different regions of the world. His research interests include broadband wireless networks, array signal processing, DSP and VLSI applications, and multimedia signal processing.

Dr. Liu has published more than 50 journal articles and has over 70 awarded patents. He is the author of "OFDM-Based Broadband Wireless Networks - Design and Optimization," Wiley 2005, and "Signal Processing Applications in CDMA Communications," Artech House, 2000. Dr. Liu's activities for the IEEE Communications Society include membership on several technical committees and serving as an editor for the IEEE Transactions in Communications. He was selected Fellow of IEEE for contributions to global standards for broadband cellular and mobile broadcasting. He is the General Chairman for the 2005 Asilomar conference on Signals, Systems, and Computers. He is a recipient of 1997 National Science Foundation (NSF) CAREER Award, the Gold Prize Patent Award in China, and 2000 Office of Naval Research (ONR) Young Investigator Award.
\end{IEEEbiography}

\begin{IEEEbiography}{Zhiyong Chen}
received the Ph.D degree in the School of Information and Communication Engineering in 2011 from Beijing University of Posts and Telecommunications (BUPT). From 2009 to 2011, he was a visiting Ph.D student in the Department of Electronic Engineering at University of Washington, Seattle, USA. He is currently an Assistant Professor in the Department of Electronic Engineering, Shanghai Jiao Tong University (SJTU), Shanghai, China. His research expertise and interests include cooperative communications, physical-layer network coding (PLNC), coded modulation in 5G mobile communication systems, computing communications. Dr. Chen serves as a publicity chair for IEEE ICCC 2014 and a TPC member for major international conferences.
\end{IEEEbiography}

\begin{IEEEbiography}{Xiaohua Tian}
(S' 07-M' 14) received his B.E. and M.E. degrees in communication engineering from Northwestern Polytechnical University, Xi'an, China, in 2003 and 2006, respectively. He received the Ph.D. degree in the Department of Electrical and Computer Engineering (ECE), Illinois Institute of Technology (IIT), Chicago, in Dec. 2010. Since June 2013, he has been with Department of Electronic Engineering of Shanghai Jiao Tong University as an Assistant Professor with the title of SMC-B scholar. He serves as an editorial board member on the computer science subsection of the journal SpringerPlus, and the guest editor of International Journal of Sensor Networks. He also serves as the technical program committee (TPC) member for IEEE INFOCOM 2014-2015, best demo/poster award committee member of IEEE INFOCOM 2014, TPC co-chair for IEEE ICCC 2014 International workshop on Internet of Things 2014, TPC Co-chair for the 9th International Conference on Wireless Algorithms, Systems and Applications (WASA 2014) on Next Generation Networking Symposium, local management chair for IEEE ICCC 2014, TPC member for IEEE GLOBECOM 2011-2015 on Wireless Networking, Cognitive Radio Networks and Ad Hoc and Sensor Networks Symposium, TPC member for IEEE ICC 2013 and 2015 on Ad Hoc and Sensor Networks and Next Generation Networking Symposium, respectively.
\end{IEEEbiography}

\begin{IEEEbiography}{Xinbing Wang}
received the B.S. degree (with honors.) from the Department of Automation, Shanghai Jiao Tong University, Shanghai, China, in 1998, and the M.S. degree from the Department of Computer Science and Technology, Tsinghua University, Beijing, China, in 2001. He received the Ph.D. degree, major in the Department of electrical and Computer Engineering, minor in the Department of Mathematics, North Carolina State University, Raleigh, in 2006. Currently, he is a professor in the Department of Electronic Engineering, Shanghai Jiao Tong University, Shanghai, China. Dr. Wang has been an associate editor for IEEE/ACM Transactions on Networking and IEEE Transactions on Mobile Computing, and the member of the Technical Program Committees of several conferences including ACM MobiCom 2012,2014, ACM MobiHoc 2012-2015, IEEE INFOCOM 2009-2015.
\end{IEEEbiography}

\begin{IEEEbiography}{Meixia Tao}
(S'00-M'04-SM'10) received the B.S. degree in electronic engineering from Fudan University, Shanghai, China, in 1999, and the Ph.D. degree in electrical and electronic engineering from Hong Kong University of Science and Technology in 2003. She is currently a Professor with the Department of Electronic Engineering, Shanghai Jiao Tong University, China. Prior to that, she was a Member of Professional Staff at Hong Kong Applied Science and Technology Research Institute during 2003-2004, and a Teaching Fellow then an Assistant Professor at the Department of Electrical and Computer Engineering, National University of Singapore from 2004 to 2007. Her current research interests include cooperative communications, wireless resource allocation, MIMO techniques, and physical layer security.

Dr. Tao is an Editor for the \textsc{IEEE Transactions on Communications} and the \textsc{IEEE Wireless Communications Letters}. She was on the Editorial Board of the \textsc{IEEE Transactions on Wireless Communications} from 2007 to 2011 and the \textsc{IEEE Communications Letters} from 2009 to 2012. She also served as Guest Editor for \textsc{IEEE Communications Magazine} with feature topic on LTE-Advanced and 4G Wireless Communications in 2012, and Guest Editor for \textsc{EURISAP J WCN} with special issue on Physical Layer Network Coding for Wireless Cooperative Networks in 2010.

Dr. Tao is the recipient of the IEEE Heinrich Hertz Award for Best Communications Letters in 2013, the IEEE ComSoc Asia-Pacific Outstanding Young Researcher Award in 2009, and the International Conference on Wireless Communications and Signal Processing (WCSP) Best Paper Award in 2012.
\end{IEEEbiography}

\end{document}